\documentclass[conference]{IEEEtran}

\IEEEoverridecommandlockouts

\usepackage{graphicx}
\usepackage[utf8]{inputenc}

\graphicspath{{figures/}}

\usepackage{url}
\usepackage{wrapfig}
\usepackage{units}

\usepackage[printonlyused]{acronym}

\usepackage{calc}
\usepackage{graphicx}
\usepackage{amsmath,amssymb,amsthm,amsfonts,amsbsy}

\usepackage{algpseudocode}
\usepackage{color}
\usepackage[table]{xcolor} 
\usepackage{nicefrac}
\usepackage{txfonts}

\usepackage{multirow}
\usepackage{amsmath}  
\usepackage{xcolor}   
\usepackage{listings}
\usepackage{enumitem}

\usepackage[multiuser,nomargin,marginclue,footnote]{fixme}
\fxusetheme{color}
\fxsetup{targetlayout=colorcb}
\FXRegisterAuthor{all}{anall}{ALL}
\FXRegisterAuthor{md}{anmd}{MD}
\FXRegisterAuthor{tolja}{antolja}{TOLJA}
\FXRegisterAuthor{pg}{anpg}{PG}
\FXRegisterAuthor{mch}{anmc}{MC}
\FXRegisterAuthor{cp}{ancp}{CP}
\FXRegisterAuthor{awo}{anawo}{AWO}

\fxsetup{status=draft}

\definecolor{airforceblue}{rgb}{0.36, 0.54, 0.66}
\definecolor{awesome}{rgb}{1.0, 0.13, 0.32}

\newcommand{\proposalName}{DeepCTC}

\begin{document}

\title{Deep Learning for Cross-Technology Communication Design}

\author{
\IEEEauthorblockN{Anatolij Zubow, Piotr Gawłowicz, Suzan Bayhan}
\IEEEauthorblockA{\{zubow, gawlowicz, bayhan\}@tkn.tu-berlin.de}
Technische Universität Berlin, Germany\\
}

\maketitle


\begin{abstract}
Recently, it was shown that a communication system could be represented as a deep learning~(DL) autoencoder.
Inspired by this idea, we target the problem of OFDM-based wireless cross-technology communication~(CTC) where both in-technology and CTC transmissions take place simultaneously.
We propose \proposalName, a DL-based autoencoder approach allowing us to exploit DL for joint optimization of transmitter and receivers for both in-technology as well as CTC communication in an end-to-end manner.
Different from classical CTC designs, we can easily weight in-technology against CTC communication.
Moreover, CTC broadcasts can be efficiently realized even in the presence of heterogeneous CTC receivers with diverse OFDM technologies. 
Our numerical analysis confirms the feasibility of \proposalName~as both in-technology and CTC messages can be decoded with sufficient low block error rate.
\end{abstract}

\begin{IEEEkeywords}
Machine Learning, networking research, wireless, cross-technology communication
\end{IEEEkeywords}

%

\section{Introduction}
The latest advances in the area of wireless cross-technology communication~(CTC) opened new ways to mitigate the coexistence problems in dense deployments of heterogeneous wireless technologies in the unlicensed spectrum~\cite{chai2016lte, zubow2018practical,WeBee}.
CTC enables direct over-the-air communication between heterogeneous devices, e.g. LTE with WiFi~\cite{gawlowicz18_infocom} or WiFi with ZigBee~\cite{WeBee}, that is otherwise not possible due to incompatible communication layers.
However, designing efficient, i.e. high-data rate, and generic CTC solutions is very challenging.
Existing CTC proposals are either technology-specific and therefore costly, or generic but inefficient, hence, limiting the scope of possible coexistence applications, i.e., only medium or even only long-term cross technology management.

Machine learning~(ML) in general and deep learning~(DL) in particular is successfully applied in fields like computer vision. Despite being extremely challenging to describe images from the real world with traditional mathematical models for detection and classification of objects in images, an ML/DL based algorithm can easily recognize images given sufficient number of training examples. 
Similarly, efficient CTC is hard to realize using conventional rule-based design due to several reasons.
First, due to the diversity of technologies and their incompatible communication layers, CTC solutions need to be designed for each pair of supported technology separately which is very costly as with $N$ different technologies ${N \choose 2}$ CTC solutions need to be developed.
Second, due to spectrum diversity even two nodes of same technology may require different CTC signals as they may use only partially overlapping spectrum bands, e.g. WiFi on two overlapping channels.
Third, as the CTC signal is transmitted overlay, i.e. superimposed on the underlying in-technology transmission, it may result in mutual interference which need to be carefully controlled as otherwise certain QoS assertion cannot be met for the in-technology communication.
Forth, in an environment with a myriad of diverse technologies, the support of an efficient CTC broadcast operation is crucial.
Instead of transmitting the CTC data to each CTC receiver independently, e.g. in round-robin fashion, the direct utilization of the broadcast nature of the wireless channel is more efficient but requires the design of a CTC signal which can be decoded by all the heterogeneous receivers in vicinity.
As today's available models fall short of capturing all aspects of CTC, we expect ML/DL-based design to yield significant improvements in CTC.
%
As recently shown in the literature, e.g. by O’Shea et al.~\cite{o2017introduction},  ML/DL-based design is promising as even a complicated communication system can be represented as a DL autoencoder.
Moreover, it becomes practically relevant to use ML/DL as the computational resources of communication devices grow substantially and first ML-specialized ASIC technology  has already become commercially available, e.g. Google TPU or Nvidia Jetson.

\smallskip
\noindent
\textbf{Contributions:} Our key contributions are as follows:
\begin{itemize}[noitemsep, topsep=0pt,leftmargin=*]
    \item With propose \proposalName~and show that it is possible to learn full transmitter and receiver implementations in an OFDM-based communication network with in-technology and CTC communication competing for capacity. This is challenging as at the transmitter side the CTC signal needs to be superimposed on the in-technology signal resulting in mutual interference. Such a setup can be represented as a deep neural network~(DNN) with two inputs and multiple outputs, where one output represents the decoded message at the in-technology receiver and the others are the decoded CTC message at each CTC receiver. The resulting DNN can be trained as an autoencoder. The learned transmitter and receiver implementations are jointly optimized towards certain performance metrics like block error rate (BLER).
    \item \proposalName~allows us to arbitrarily weight the in-technology against CTC communication, i.e. to achieve prioritization in terms of BLER.
    \item \proposalName~allows a transmitter to learn optimal CTC broadcasting strategy, which is paramount especially in coexistence settings with many CTC receivers with diverse OFDM technologies, e.g. LTE and WiFi. 
\end{itemize}

%

\section{Related Work}
Recently, the usage of ML/DL techniques for physical layer design of communication systems sparked the interest of the research community.
O’Shea et al.~\cite{o2017introduction} have shown that a communication system can be represented as a DL autoencoder.
Hence, the communications system design can be seen as an end-to-end reconstruction task that seeks to jointly optimize transmitter and receiver components in a single process.
The authors showed that the idea can be extended to networks of multiple transmitters and receivers, i.e. interference channel.
In~\cite{o2017deep}, this idea was extended to a single-user end-to-end MIMO communication system over a Rayleigh fading channel.
Moreover, Borgerding et al.~\cite{borgerding2017amp} showed that DL can be used to recover the sparse signal from noisy linear measurements in MIMO environments.
%
Here the proposed scheme was able to outperform the traditional algorithms in estimating the massive-MIMO channel.
Ye et al.~\cite{ye2018power} showed how DL can be used for channel estimation and signal detection in OFDM systems.
In this paper, we extend the related work towards an OFDM-based CTC system with in-technology and CTC transmissions.
%

%

\section{Background Knowledge}
This section provides a brief overview of OFDM and DL.
\subsection{OFDM Primer}\label{chapter:ofdm_primer}
%
%
Due to its high spectral efficiency and robustness to frequency-selective fading, many wireless technologies, e.g., 3GPP LTE and IEEE WiFi, use Orthogonal Frequency Division Multiplexing~(OFDM).  
Moreover, it is likely that OFDM will remain the technology of choice for future wireless networks.
The idea behind OFDM is threefold: (i) parallel transmissions over multiple narrow-band channels, (ii) orthogonal subcarriers and (iii) subcarrier level modulation and coding scheme~(MCS) adaptation. 
First, an OFDM transmitter~(TX) splits a high-rate data stream into many lower-rate parallel streams.
Owing to a lower rate, the symbol duration can be longer compared to the single channel modulation schemes, which helps to reduce Inter-Symbol Interference (ISI) occurring in multipath environments.
Second, the TX divides the wireless channel bandwidth $B$ into $N$ narrow-band subcarriers such that subcarriers are orthogonal to each other and thereby transmissions on these carriers do not interfere.
Each data stream then is transmitted on one of these subcarriers.
Third, the TX can apply a different MCS for each subcarrier to cope with the unfavorable channel conditions including frequency selective fading and narrow-band interference.
%

OFDM can be efficiently realized using Fast Fourier Transform (FFT). 
Specifically, a transmitter modulates incoming data bits into subcarriers. Then, it relies on inverse FFT to convert the frequency domain representation into the time domain and sends the signal over the air interface. 
An OFDM receiver reverses the steps performed at the transmitter, thus, it executes FFT on the received signal to convert it back to the frequency domain and demodulates subcarriers independently.

%
%
%
%
%
%
The operation of an OFDM transmitter can be abstracted as spreading data on a two-dimensional grid~(in frequency (subcarrier) and time (symbol) domains) which we will refer to as \textit{OFDM time-frequency grid}~(OTFG) hereafter.
Despite relying on the same modulation approach, current OFDM-based wireless networks use vastly diverse parameters for their operation.
%
%
Table~\ref{ofdm_examples} presents typical parameters of various OFDM-based wireless technologies.

\begin{table}
\centering
\caption{List of example OFDM-based wireless technologies.}
\label{ofdm_examples}
\scriptsize
\begin{tabular}{|c|c|c|c|c|c|} 
\cline{1-6}
\begin{tabular}[c]{@{}c@{}}Wireless\\Technology \end{tabular} & \begin{tabular}[c]{@{}c@{}}Channel \\Bandwidth\\ $B$ [MHz]\end{tabular} & \begin{tabular}[c]{@{}c@{}}Sampling\\Rate \\ $f_{sr}$[MHz] \\ \end{tabular} & \begin{tabular}[c]{@{}c@{}}FFT\\size $N$\end{tabular} & \begin{tabular}[c]{@{}c@{}}Subcarrier\\Spacing \\$\Delta f_{sc}$[kHz] \end{tabular} & \begin{tabular}[c]{@{}c@{}}Symbol\\Duration \\$t_{s}$[$\mu$s] \end{tabular}  \\ 
\cline{1-6}
802.11n/ac                                                    & 17.5                                                                     & 20                                                                                & 64                                                    & 312.5                                                                                                     & 3.2                                                                                  \\
802.11ax                                                      & 17.5                                                                     & 20                                                                                & 256                                                   & 78.125                                                                                                    & 12.8                                                                                 \\
LTE-LAA/U                                                     & 18                                                                       & 30.72                                                                             & 2048                                                  & 15                                                                                                        & 66.6                                                                                 \\
WiMAX                                                         & 18.4                                                                       & 22.4                                                                              & 2048                                                  & 10.94                                                                                                     & 91.4                                                                                 \\
\cline{1-6}
\end{tabular}
\end{table}
\subsection{Deep Learning (DL) Primer}

\begin{figure}
\centering
\vspace{-10pt}
\includegraphics[width=0.8\linewidth]{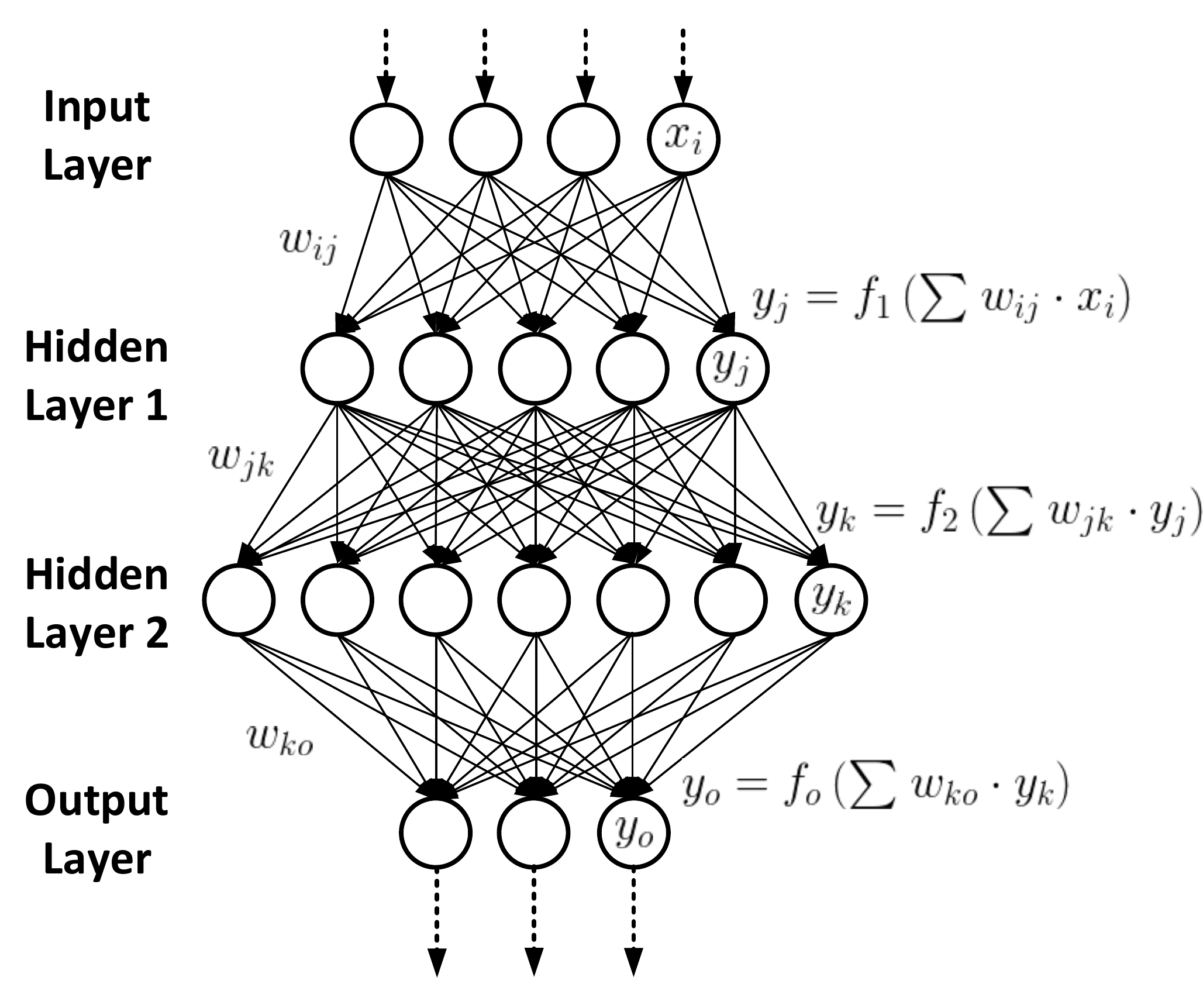}
\vspace{-5pt}
\caption{A fully connected feedforward NN architecture where all neurons
between adjacent layers are fully connected. The DNN uses multiple hidden
layers between the input and output layers to extract meaningful features.}
\label{fig:neural-networks-layers}
\vspace{-10pt}
\end{figure}

Deep learning (DL) has been successfully applied in a wide range of areas with significant performance improvement, ranging from computer vision~\cite{krizhevsky2012imagenet}, natural language processing~\cite{collobert2011natural} to speech recognition~\cite{hinton2012deep}.
The structure of the DNN architecture is given in Fig.~\ref{fig:neural-networks-layers}. 
In order to improve the ability in representation or recognition deep neural networks (DNN) use an increased
number of hidden layers.
Each network layer consists of multiple neurons, each of which has an output that is a nonlinear function of a weighted sum of neurons of its preceding layer.
The widely used nonlinear functions are Sigmoid, tanh or ReLU.
Hence, the output of the DNN $\textbf{n}$ is a cascade of nonlinear
transformation of input data \textbf{I}, mathematically expressed as~\cite{ye2018power}:
$\textbf{n} = f(\textbf{I}, \theta) = f^{L-1}(f^{L-2}(\ldots f^1(\textbf{I})))$
where $L$ is the number of layers and $\theta$ denotes the weights of the DNN.
The parameters of the model are the weights of the neurons, which need to be optimized
before the network can be deployed.
Usually, the optimal weights are learned on a training set, with known desired outputs.

Numerous DL tools are available that make it simple to build, train and use large NNs.
Therefore, normally high-level programming languages like Python are used to hide the complexity of training routines using massively parallel GPU architecture.
The most widely used ML tools are Caffe and TensorFlow\footnote{https://www.tensorflow.org/}.
Keras\footnote{https://keras.io/} provides a high-level API allowing to create and optimize even a very complex NN in just a few lines of code.
Keras runs on top of TensorFlow that allows to represent numerical computation as a data flow graphs, i.e. nodes represent mathematical operations, while edges represent the multidimensional data arrays that flow between them.
It is worth to mention, that with only a minor change of configuration parameters, the TensorFlow library is able to execute the same computing graphs on a single CPU and GPU as well as distributed clusters of them.

%

\section{Cross-Technology Communication (CTC)}\label{chapter:ctc}
\begin{figure}
\centering
\includegraphics[width=0.90\linewidth]{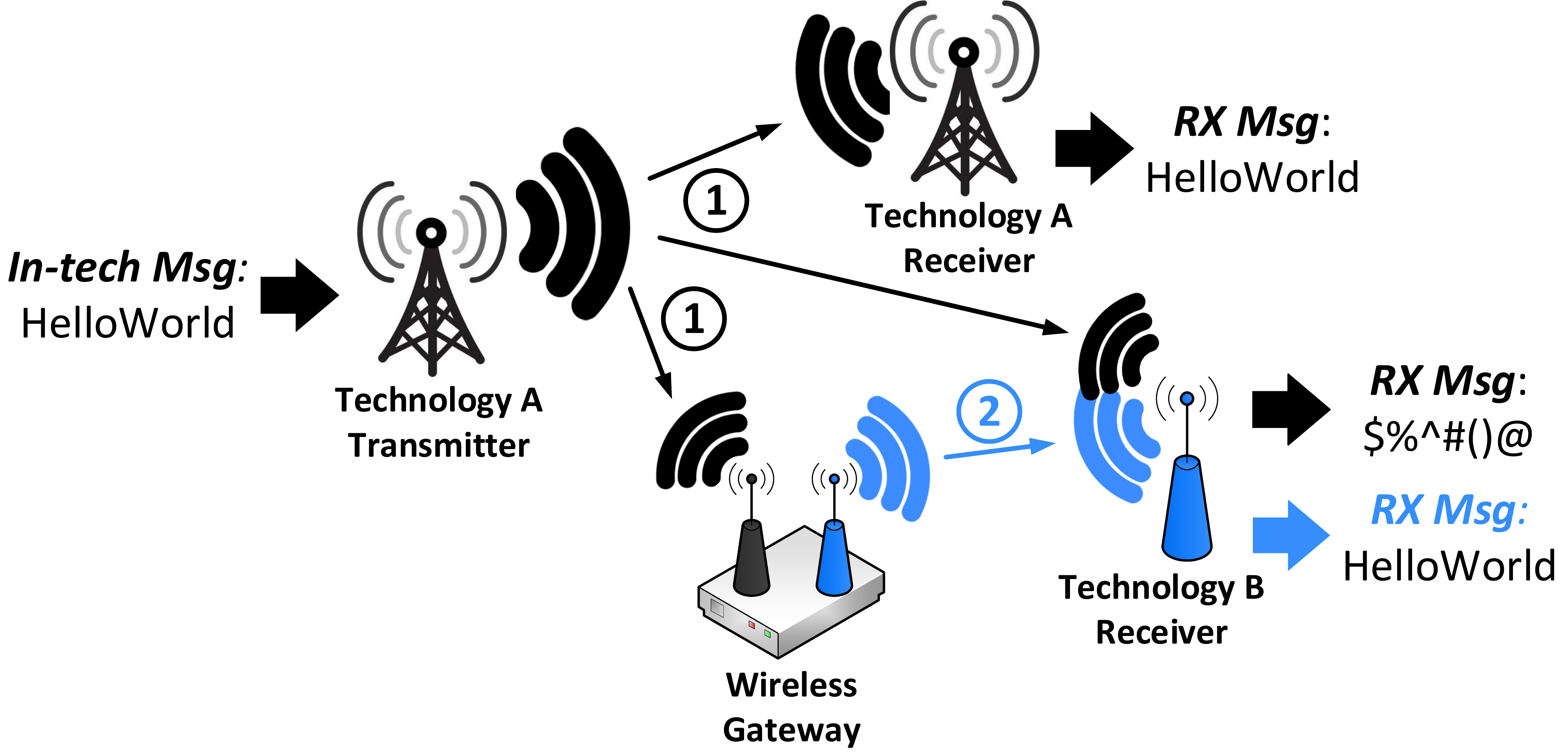}
\caption{Heterogeneous devices cannot directly communicate. Instead, they have to use a wireless gateway.}
\label{fig:autoCTC_1}
\end{figure}

\begin{figure}
\centering
\vspace{-5pt}
\includegraphics[width=0.90\linewidth]{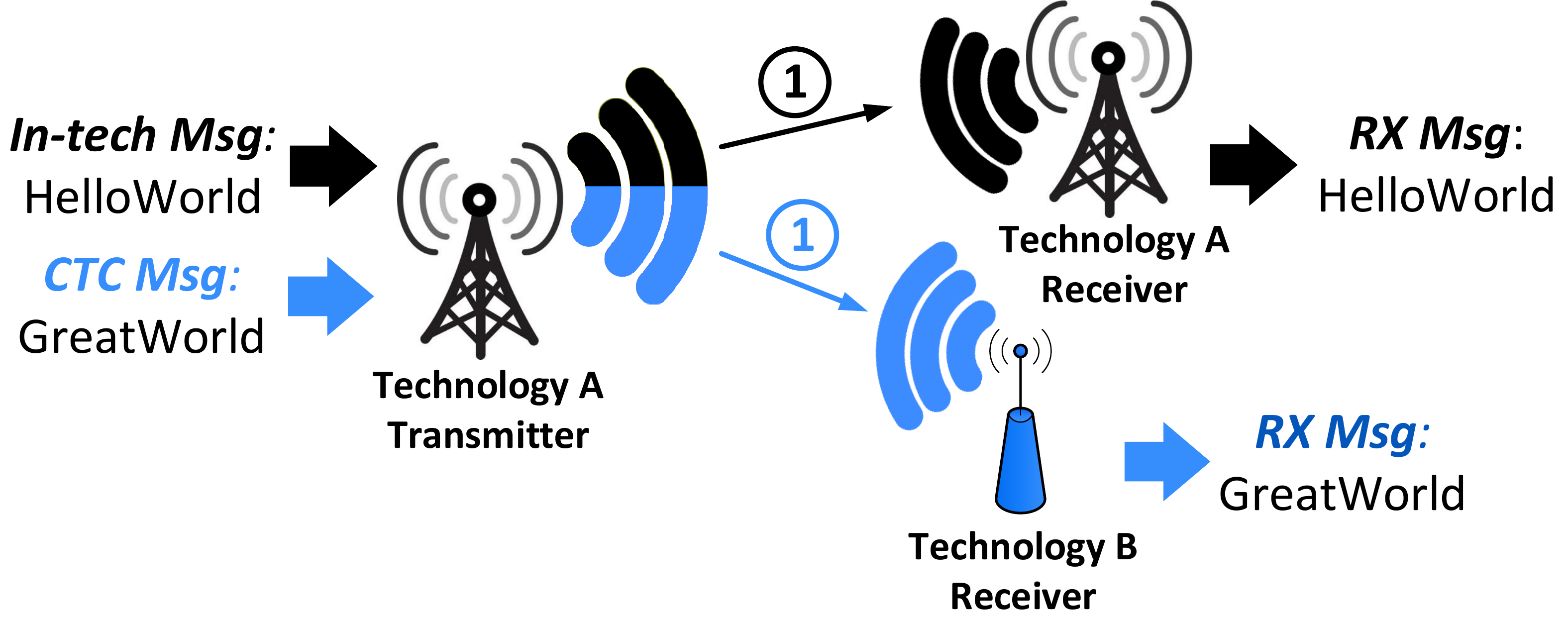}
\caption{CTC enables heterogeneous devices to talk directly.}
\vspace{-10pt}
\label{fig:autoCTC_2}
\end{figure}

The traditional way to achieve bridging between different wireless technologies is to use multi-radio gateways.
Such an indirect approach requires additional hardware for the gateway which is not only costly but also unnecessarily increases deployment complexity. Moreover, it is inefficient as the same message has to be sent multiple times, i.e. once on each technology, and may result in a bottleneck at the gateway.
As a result, a new field of research termed as CTC has emerged.
Pioneering works like~\cite{Esense,TransparentCTC} showed that it is possible to exchange messages among heterogeneous wireless technologies despite their incompatible physical layer modulation - cmp. Fig~\ref{fig:autoCTC_1} with Fig~\ref{fig:autoCTC_2}.
Known examples from literature are WEBee~\cite{WeBee} which allows a WiFi device to talk directly to a ZigBee node and LtFi~\cite{gawlowicz18_infocom} which enables CTC between LTE and WiFi.
We can categorize the prior CTC solutions into two as packet-level and physical-layer CTC solutions.
%
Packet-level solutions
convey the CTC message by modulating bits into either frame length~\cite{Esense,HoWiES}, gap or inter-frame spacing~\cite{TransparentCTC,gawlowicz18_infocom}, packet transmission power~\cite{wizig}, timing of periodic beacon interval~\cite{kim2017free}, prepending legacy packets with a customized preamble containing sequences of energy pulse~\cite{GapSense},\cite{cmorse,TransparentCTC,emf}.
On the other hand, physical-layer solutions manipulate the signal to enable CTC. Some outstanding proposals in this category are WeBee~\cite{WeBee}, TwinBee~\cite{chen2018twinbee} and LongBee~\cite{li2018longbee} which target CTC between WiFi and ZigBee.
In WeBee, a WiFi device emulates the ZigBee OQPSK signal by properly selecting the payload of WiFi frame.
In a similar spirit, ULTRON~\cite{chai2016lte} proposes to emulate a WiFi OFDM signal at the LTE-U transmitter.

\begin{figure}[!ht]
\centering
\vspace{-5pt}
\includegraphics[width=1.00\linewidth]{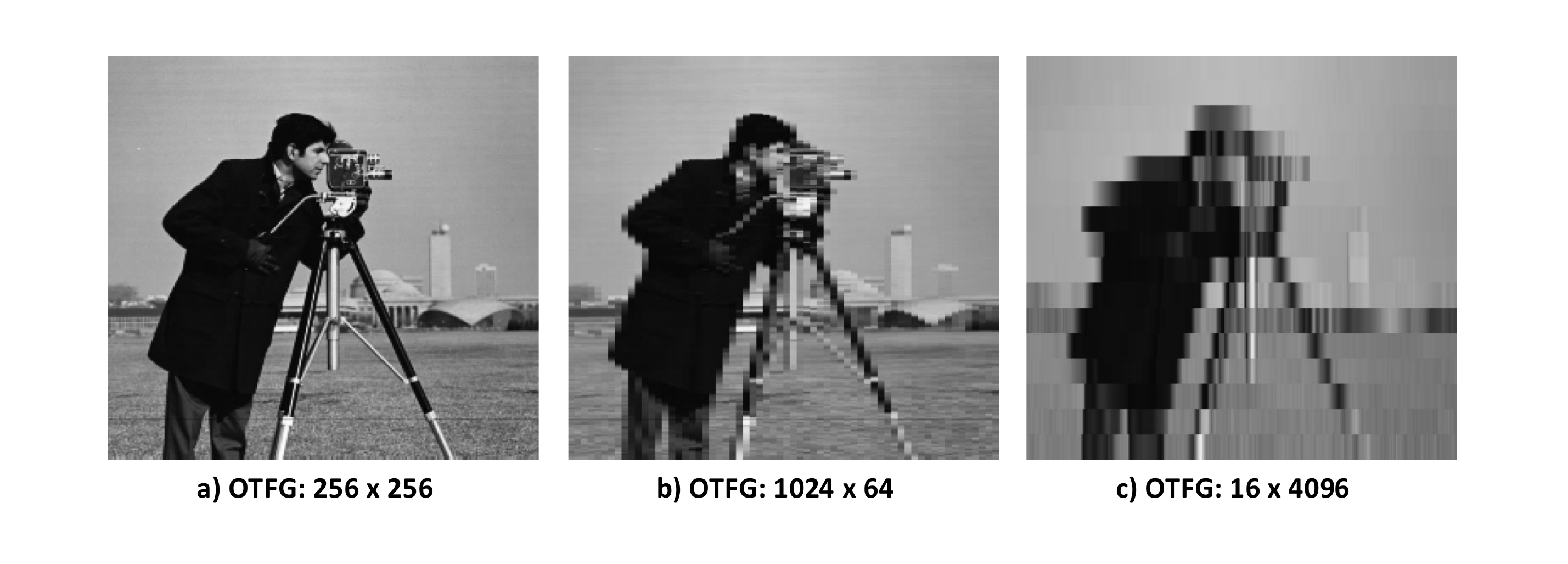}
\vspace{-15pt}
\caption{Image processing (i.e. grid resampling) as the analogy of cross-technology OFDM signal reception --- due to different OTFG dimensions at transmitter (a) and CTC receivers (b and c), the transmitted signal cannot be fully recovered at the receivers.}
\vspace{-5pt}
\label{fig:ctc_illu}
\end{figure}

In this paper, we focus on the problem of joint encoding of in-technology and CTC data for OFDM-based wireless technologies.
Note that such a CTC modulation where the CTC signal is superimposed on top of the in-technology communication can be seen as a form of downlink superposition coding~\cite{tse2005fundamentals}.
%
%
Note as different OTFGs are used at transmitter and CTC receiver, the cross-technology OFDM signal reception is not trivial.
Specifically, orthogonal OFDM symbols of the transmitter interfere with each other in time and/or frequency in the heterogeneous receiver.
As an example, consider the case of LTE and WiFi~(802.11n) where an LTE OFDM symbol is $\approx 21 \times$ longer in time and $\approx 21 \times$ narrower in frequency as compared to WiFi~(Table~\ref{ofdm_examples}).
In Fig.~\ref{fig:ctc_illu}, we illustrate the issue of cross-technology OFDM signal reception using an analogy from the image processing area.

%

\section{Design Goals and Overview of \proposalName}
%
As mentioned in §~\ref{chapter:ctc}, our focus is on CTC modulations where the CTC message is superimposed on top of the underlying in-technology transmission.
Such an approach creates an overhead on the underlying in-technology transmission either in increased bit error rate due to mutual interference or increased frame length as additional redundancy for forward error correction might be needed.
Hence, such a trade-off needs to be considered when designing a CTC solution.

The second challenge is the fact that in a mixed environment a node is surrounded by different types of OFDM technologies which differ in their configuration, mainly in their OTFG, i.e. sizes of the FFT bins in both time and frequency.
Hence, the transmitter has to account for that mismatch when encoding CTC messages for the receiver.
This is especially challenging as for an efficient CTC broadcast, i.e. a CTC message needs to be received and decoded by all CTC receivers having different OTFG. 
Hence, we envision the following ML/DL approach.
First, a node joining the radio spectrum performs scanning for different technologies in vicinity using methods like~\cite{rajendran2018deep,olbrich2017wiplus}.
After discovering the OTFG dimensions used by the technologies in vicinity, the node loads the corresponding pre-trained \proposalName~model from which it automatically derives the transmitter and receiver functionality.
When a technology leaves the vicinity, the aforementioned process needs to be repeated by excluding the leaving technology.
Note that in typical coexistence scenarios usually only the base stations of the networks, e.g., LTE eNB and WiFi AP, need to directly communicate for coordination of their spectrum sharing.
Hence, as those entities are usually static and there is only a limited number of available technologies, it is feasible to preload the trained models on the devices.

%

\section{System Model}

\begin{figure}[!ht]
\centering
\vspace{-5pt}
\includegraphics[width=1.00\linewidth]{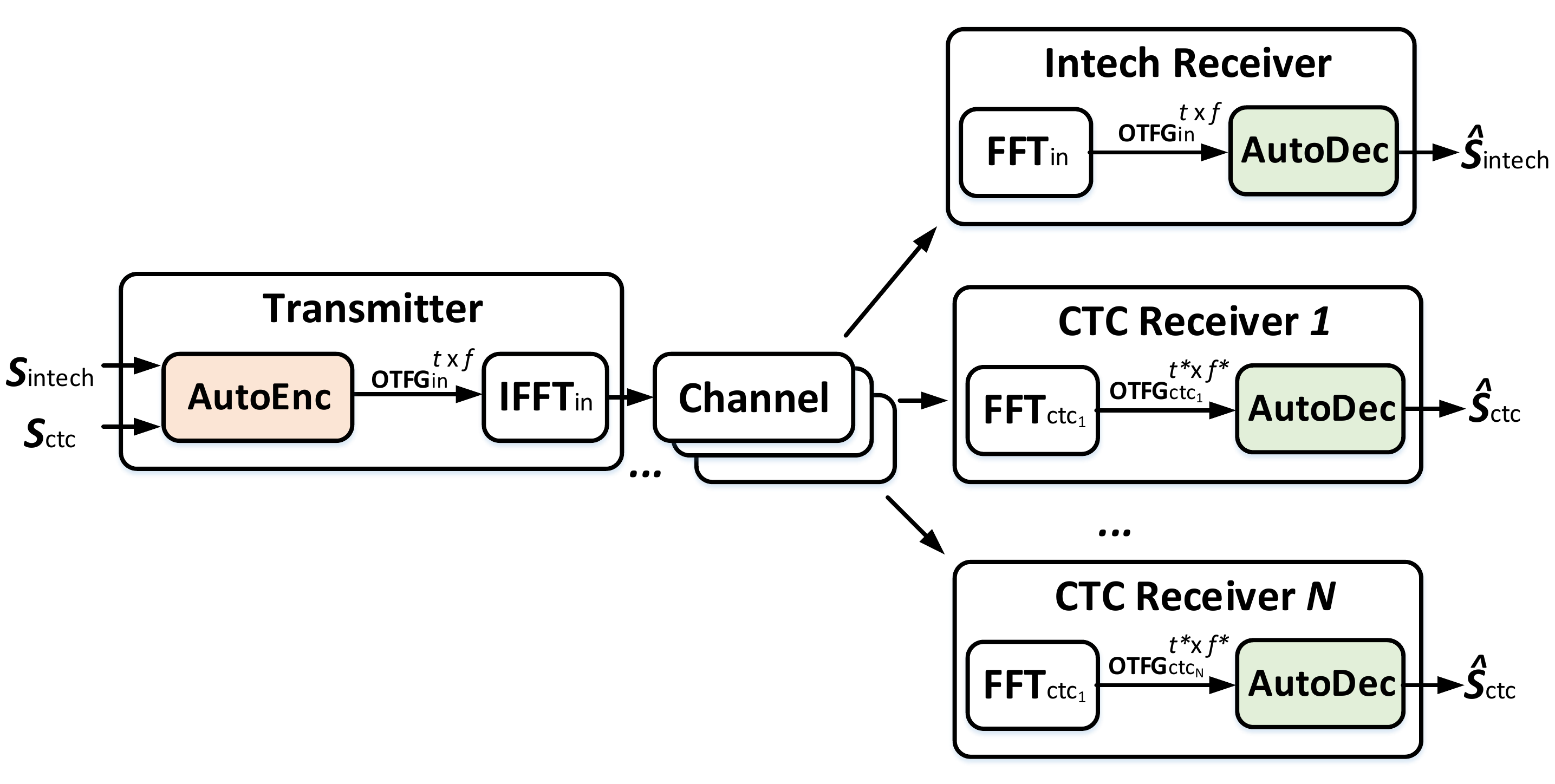}
\vspace{-20pt}
\caption{System model with in-technology and CTC (broadcast) communication consisting of a transmitter, an in-technology receiver and multiple CTC receivers connected through a channel.}
\label{fig:ctc_system}
\vspace{-10pt}
\end{figure}

An OFDM-based communication system with in-technology and CTC communication is shown in Fig.~\ref{fig:ctc_system}.
It consists of a transmitter, channel(s) and multiple receivers. Moreover the receivers can be classified into in-technology and CTC receiver(s).
The transmitter wants to communicate two types of messages: i) in-technology message $s_{intech}$ and ii) CTC message $s_{ctc}$ to the in-technology and CTC receivers, respectively.
Therefore, it selects one out of $M$ possible messages $s_{intech} \in \mathcal{M} = \{1,2,\ldots,M\}$ for in-technology and also one out of $C$ possible messages $s_{ctc} \in \mathcal{C} = \{1,2,\ldots,C\}$ for CTC communication.
The two messages are superimposed on each other, mapped to the transmitters OTFG and communicated to the receiver(s) making $t \times f$ discrete uses of the channel\footnote{$t \times f$ is the size of the transmitter's OTFG.}.
Hence the transmitter applies the transformation $f_{intech}: \mathcal{M} \rightarrow \mathbb{R}^{t \times f}$ to the message $s_{intech}$ and transformation $f_{ctc}: \mathcal{C} \rightarrow \mathbb{R}^{t \times f}$ to the message $s_{ctc}$ to generate the transmitted signal $\textbf{x} = f_{intech}(s_{intech}) + f_{ctc}(s_{ctc})$.
In addition the transmitter imposes certain constraints on $\textbf{x}$ like energy, power and amplitude constraints~\cite{o2017introduction}.

The communication rate of such a communication system is $R = k/(tf)$ [bit per OTFG use], where $k = log_2{M}$, and $R = j/(tf)$, where $j = log_2{C}$, for in-technology and CTC communication, respectively.
The channels are described as conditional PDFs $p(\textbf{y}_{intech}|\textbf{x})$, where $\textbf{y}_{intech} \in \mathbb{R}^{t \times f}$ denotes the received signal at the in-technology receiver.
The signal for the CTC receivers passes the channel $p(\textbf{y}_{ctc}|\textbf{x})$, where $\textbf{y}_{ctc} \in \mathbb{R}^{t^* \times f^*}$ denotes the received signal at CTC receiver.
Note that as the CTC receivers uses a different OTFG, $\textbf{y}_{ctc}$ has a different dimension as $\textbf{x}$.
Upon reception, the in-technology receiver applies the transformation $g_{intech}: \mathbb{R}^{t \times f} \rightarrow \mathcal{M}$ to produce the estimate $\hat{s}_{intech}$ of the transmitted message $s_{intech}$.
A similar transformation is applied at the CTC receivers, i.e. $g_{ctc}: \mathbb{R}^{t^* \times f^*} \rightarrow \mathcal{C}$ to produce the estimate $\hat{s}_{ctc}$ of the transmitted message $s_{ctc}$.
Recall that CTC receivers might use different values for $t^*$ and $f^*$, i.e., different OTFG.

From the perspective of DL, such a communication system can be seen as a type of autoencoder~\cite{liou2014autoencoder}.
The task of the autoencoder in the context of a communication system is to learn representations $\textbf{x}$ of the transmitted messages that are robust with
respect to the channel impairments mapping $\textbf{x}$ to $\textbf{y}$ (i.e., noise,
fading, distortion, etc.), so that the transmitted message can be recovered with small probability of error~\cite{o2017introduction}. 
%

In the case of CTC, the autoencoder is even more different.
In addition to learning an intermediate representation robust to channel impairments, it has to learn how to superimpose the CTC signal on top of in-technology signal robust to mutual interference.
Moreover, the autoencoder needs to learn that the in-technology and CTC receivers are diverse as they use OTFGs of different dimensions.
Hence it has to learn an intermediate representation that allows both the in-technology and the CTC receivers to recover their respective messages with small probability of error.

Note that the proposed approach can be extended to CTC systems where the nodes use different but at least partially overlapping spectrum, e.g. two WiFi nodes operating on overlapping channels in 2.4\,GHz band.
Here, only the overlapping spectrum can be used for transmission of CTC messages.

%

\section{\proposalName: An Autoencoder for CTC}

\subsection{Model Building}
\begin{figure}
\centering
\vspace{-5pt}
\includegraphics[width=0.56\linewidth]{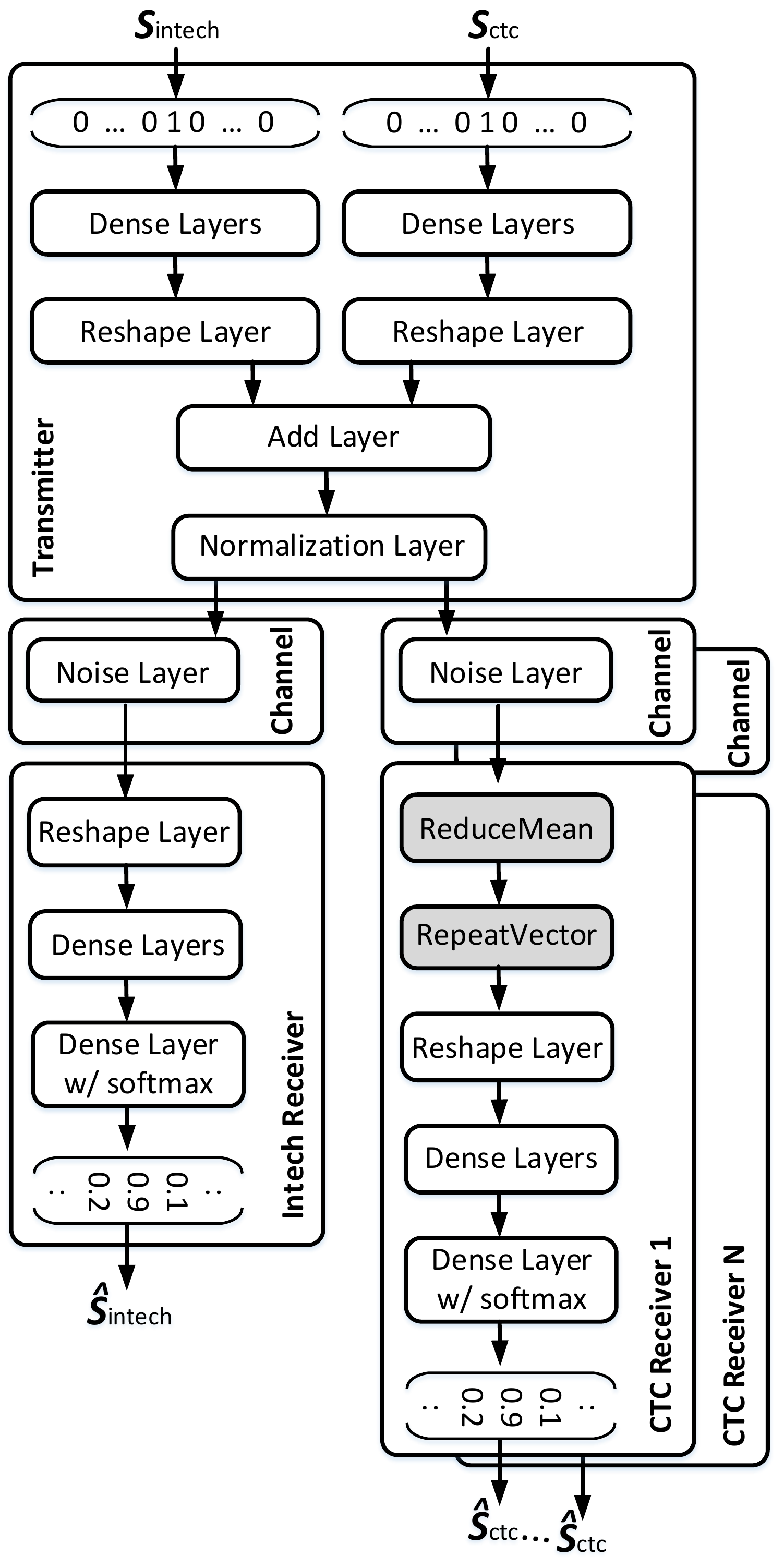}
\vspace{-5pt}
\caption{A communication system with in-technology and CTC transmission over an AWGN channel represented as DL autoencoder.}
\vspace{-5pt}
\label{fig:ctc_autoencoder}
\end{figure}

The proposed autoencoder for CTC is shown in Fig.~\ref{fig:ctc_autoencoder}.
The transmitter consists of a feedforward NN with multiple dense layers and a reshape layer for each input message, in-technology and CTC.
The reshaping is needed in order to map the signal to the OTFG used by the transmitter.
The outputs of the two reshape layers is merged using an add layer, i.e. superposition of the in-technology and CTC messages.
The final layer is the normalization layer that ensures that the physical constraints on $\textbf{x}$ are met.
Note that the two inputs $s_{intech}$ and $s_{ctc}$ to the transmitter are encoded as a one-hot vector.
The OFDM channel is represented by a noise layer with fixed variance.
Both the in-technology and CTC receivers are implemented as feedforward NNs.

The in-technology receiver is simpler. The first layer is a reshape layer which converts the parallel OFDM signal into a serial one.
It is followed by multiple dense layers whereas the last one uses softmax activation whose output $\textbf{p} \in (0,1)^M$ is a probability vector over all possible in-technology messages.
Here the decoded message $\hat{s}_{intech}$ corresponds to the index of the element of $\textbf{p}$ with the highest probability.
The CTC receiver is slightly different as it contains two additional layers before the reshape layer, namely reducemean and repeatvector layers.
Those two layers are needed to map transmitter's OTFG on the one used by the CTC receiver, i.e. different OFDM parameters like FFT size.

\subsection{Model Training}
The models are trained by viewing OFDM (de-)modulation, i.e. (I)FFT, and the wireless channels as black boxes (see Fig.~\ref{fig:ctc_system}).
The autoencoder can be trained end-to-end using stochastic gradient descent (SGD) on a set of all possible in-technology $s_{intech} \in \mathcal{M}$ and CTC $s_{ctc} \in \mathcal{C}$ messages.
Therefore, we take the individual cross-entropy loss functions at the in-technology and
CTC receiver, respectively.
We train the two coupled autoencoders with conflicting goals.
Our approach consists of minimizing a weighted sum of both losses, i.e., $L = \alpha L_{intech} + (1 - \alpha) L_{ctc}$ for some $\alpha \in [0,1]$.
This is needed as minimizing $L_{intech}$ alone ($\alpha = 1$) would result in the transmitter to transmit a weak and/or constant signal for $s_{ctc}$ that the in-technology receiver can simply subtract from $\textbf{y}_{intech}$. However, it would make the decoding of $s_{ctc}$ at the CTC receiver impossible.
By selecting a proper $\alpha$, a proper weight can be assigned to in-technology and CTC transmission, hence allowing to control the trade-off between the decodability~(BLER) of the two transmissions.

%

\section{Implementation}
We have implemented the proposed CTC autoencoder using open source ML libraries like Tensorflow and Keras.
Most of the needed NN layers are part of keras like \texttt{Input}, \texttt{Dense}, \texttt{GaussianNoise}, \texttt{Reshape}, \texttt{RepeatVector} and \texttt{Add}.
Missing layers like \texttt{ReduceMean} where implemented using keras \texttt{Lambda} layer and Tensorflow.
The model was trained with a data set of size $10^6$ whereas the testing data size was $2 \times 10^6$.
After the training the autoencoder in AWGN channel we used the trained model to derive the the encoder used by the transmitter and the two decoders used by the in-technology and CTC receivers respectively.
%

%

\section{Results}
We have performed several experiments to demonstrate the feasibility of DL-based CTC design.
This section summarizes the results obtained from numerical simulations.
\subsection{Joint transmission of in-technology and CTC traffic}

\begin{figure}
\centering
\vspace{-5pt}
\includegraphics[width=0.95\linewidth]{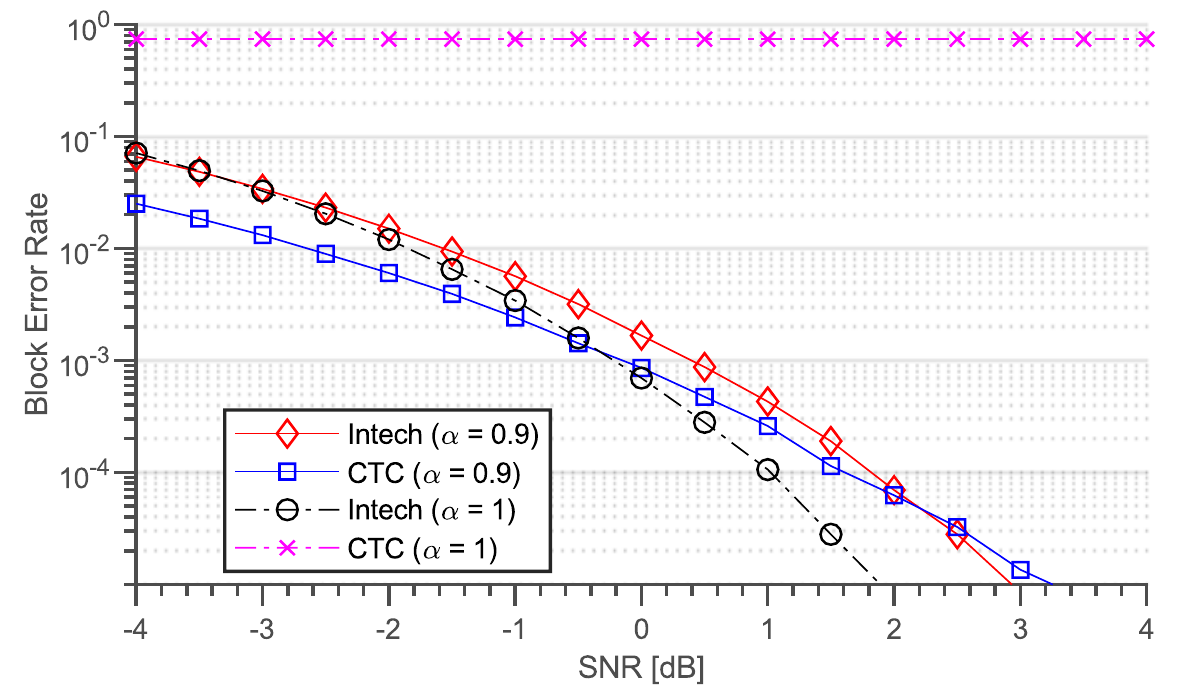}
\vspace{-10pt}
\caption{BLER of in-technology and CTC communication for different $\alpha$.}
\vspace{-5pt}
\label{fig:ctc_joint_ber}
\end{figure}

The proposed approach allows to weight the importance of in-technology communication against the superimposed CTC by controlling the value of $\alpha$.
Fig.~\ref{fig:ctc_joint_ber} shows the block error rate (BLER), i.e., $Pr(\hat{s}_{intech} \neq s_{intech})$ and $Pr(\hat{s}_{ctc} \neq s_{ctc})$, of a communications system employing binary phase-shift keying (BPSK) modulation, $|\mathcal{M}| = 64$, $|\mathcal{C}| = 4$ and an OTFG on the transmitter side of size $4 \times 4$.
While the in-technology receiver uses the same 2D grid as the transmitter, the CTC receiver's grid was $16 \times 1$, i.e. FFT size was $4 \times$ larger.

We can observe in Fig.~\ref{fig:ctc_joint_ber} that the CTC transmission fails with probability of 1 even for high SNR if $\alpha=1$.
This is because the full preference is given to the in-technology communication.
With a slightly smaller value $\alpha=0.9$, both communications can proceed simultaneously. 
Note the slight degradation of the in-technology communication of at most 1\,dB.
For $\alpha=0.9$ and SNR of $\approx 3$\,dB, our \proposalName~achieves almost error-free communication for both CTC and in-technology traffic.

\subsection{Serving CTC broadcast}

In case of multiple heterogeneous CTC receivers, efficient CTC broadcasting is paramount as otherwise a CTC transmitter has to send its CTC message to each CTC receiver in CTC unicast mode in a round-robin fashion. 
%
%
Instead, we can let the proposed autoencoder to learn efficient broadcast communication towards multiple heterogeneous CTC receivers.
Therefore, in this experiment, we disable in-technology communication and send only CTC messages towards two CTC receivers having different FFT sizes, one with $4 \times 4$ OTFG and the other with $16 \times 1$ OTFG.
%
The autoencoder was configured with a weight of 1 for both receivers and trained.

\begin{figure}
\centering
\vspace{-5pt}
\includegraphics[width=0.95\linewidth]{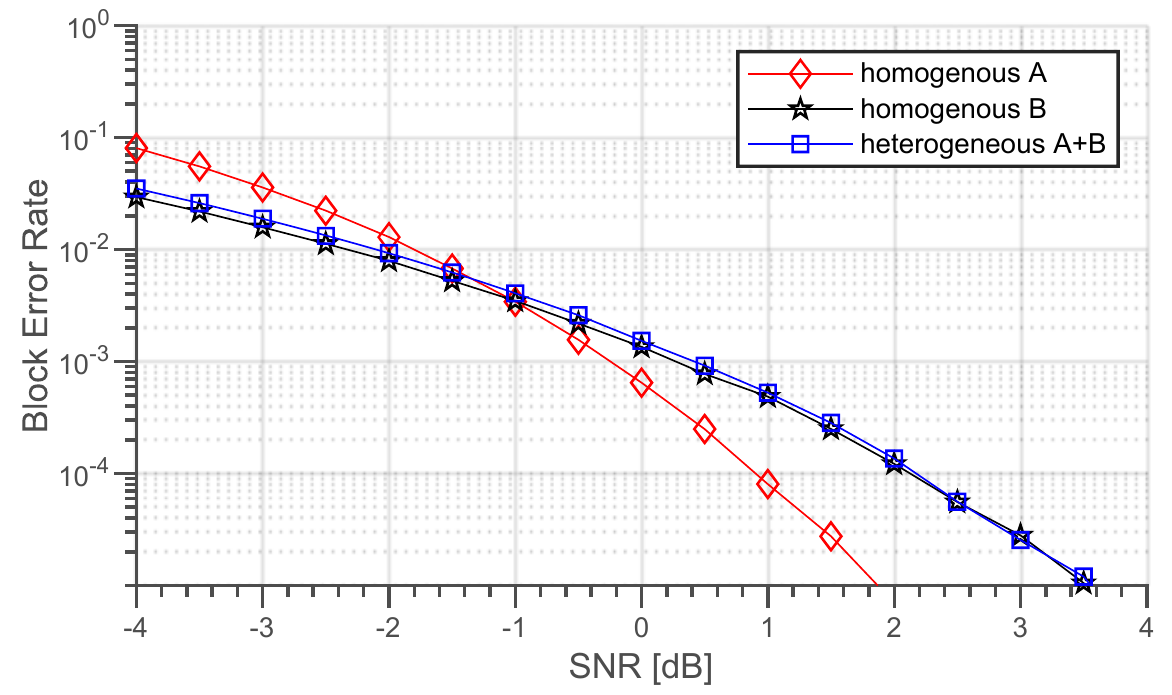}
\vspace{-10pt}
\caption{BLER of CTC broadcast communication towards a group of 2 nodes. }
\vspace{-5pt}
\label{fig:ctc_bcast}
\end{figure}

Fig.~\ref{fig:ctc_bcast} shows the results.
We see three curves for the three considered scenarios. 
Two curves represent the scenarios with homogeneous receivers, i.e. both receivers are configured with either $4 \times 4$ or $16 \times 1$ OTFG.
The third curve represents the mixed environment where one receiver uses $4 \times 4$ OTFG whereas the other $16 \times 1$ OTFG.
As we are interested in the broadcast performance, each curve shows maximum of the individual BLERs, i.e. $\max{(\mathrm{BLER}_1, \mathrm{BLER}_2)}$ where $\mathrm{BLER}_1$ and $\mathrm{BLER}_2$ are the BLER experienced by the two CTC receivers.
We can see that \proposalName~is able to learn to encode the CTC signal in such a way that it can be decoded with high probability by two CTC receivers with different OTFG sizes - cmp. curves \textit{homogeneous B} with \textit{heterogeneous A+B}.

%

\section{Conclusions \& Future Work}

This paper shows that a cross-technology communication (CTC) system can be represented as a deep learning~(DL) autoencoder.
This approach allows us on one hand to jointly optimize the transmitter and receiver of both in-technology and CTC communication.
On the other hand, efficient CTC broadcasting becomes possible.
As future work, we plan to perform end-to-end learning over real channels and hardware.
We expect DL to give significant improvements as technology, hardware, and channel together form a black-box whose input and output can be observed, but for which no exact analytic expression is known a priori~\cite{o2017introduction}.
Another goal is to perform transfer learning where the E2E CTC system trained on a statistical model is adopted to a real-world implementation~\cite{zhang2019deep}.

\bibliographystyle{IEEEtran}

\bibliography{biblio,IEEEabrv}

\end{document}